\documentclass[aps,prl,twocolumn,superscriptaddress,floatfix]{revtex4-2}
\usepackage{amsmath,amssymb,amsfonts,braket,graphicx,textgreek,xcolor,mathtools,bm,xfrac,footmisc,paralist}
\usepackage[colorlinks=true,allcolors=blue]{hyperref}
\usepackage[version=4]{mhchem}
\usepackage[T1]{fontenc} 
\usepackage[varg]{txfonts}

\newcommand{\mi}{\mathrm{i}}

\newcommand{\md}{\mathrm{d}}
\newcommand{\Hc}{\text{H.c.}}

\def\equationautorefname~#1\null{Eq.~(#1)\null}
\allowdisplaybreaks

\begin{document}
    \title{Chiral-Flux-Phase-Based Topological Superconductivity in Kagome Systems with Mixed Edge Chiralities}
	
    \author{Junjie Zeng}
    \affiliation{Institute for Structure and Function \& Department of Physics \& Chongqing Key Laboratory for Strongly Coupled Physics, Chongqing University, Chongqing 400044, P. R. China}
    \author{Qingming Li}
    \affiliation{Department of Physics, Shijiazhuang University, Shijiazhuang, Hebei 050035, China}
    \author{Xun Yang}
    \affiliation{Institute for Structure and Function \& Department of Physics \& Chongqing Key Laboratory for Strongly Coupled Physics, Chongqing University, Chongqing 400044, P. R. China}
    \author{Dong-Hui Xu}
    \email{donghuixu@cqu.edu.cn}
    \affiliation{Institute for Structure and Function \& Department of Physics \& Chongqing Key Laboratory for Strongly Coupled Physics, Chongqing University, Chongqing 400044, P. R. China}
    \author{Rui Wang}
    \email{rcwang@cqu.edu.cn}
    \affiliation{Institute for Structure and Function \& Department of Physics \& Chongqing Key Laboratory for Strongly Coupled Physics, Chongqing University, Chongqing 400044, P. R. China}
    \affiliation{Center for Computational Science and Engineering, Southern University of Science and Technology, Shenzhen 518055, P. R. China}
    \affiliation{Center of Quantum materials and devices, Chongqing University, Chongqing 400044, P. R. China}
    
    \date{\today}
	
    \begin{abstract}
        Recent studies have attracted intense attention on the quasi-2D kagome superconductors $ A\ce{V3Sb5} $ ($ A = $ K, Rb, and Cs) where the unexpected chiral flux phase (CFP) associates with the spontaneous time-reversal symmetry breaking in charge density wave (CDW) states. Here, commencing from the 2-by-2 CDW phases, we bridge the gap between topological superconductivity (TSC) and time-reversal asymmetric CFP in kagome systems. 
        Several chiral TSC states featuring distinct Chern numbers emerge for an s-wave or a d-wave superconducting pairing symmetry. Importantly, these CFP-based TSC phases possess unique gapless edge modes with mixed chiralities (i.e., both positive and negative chiralities), 
        but with the net chiralities consistent with the Bogoliubov-de Gennes Chern numbers. We further study the transport properties of a two-terminal junction, using Chern insulator or normal metal leads via atomic Green's function method with Landauer-Büttiker formalism. In both cases, the normal electron tunneling and the crossed Andreev reflection oscillate as the chemical potential changes, but together contribute to plateau transmissions (1 and \sfrac{3}{2}, respectively). These behaviors can be regarded as the signature of a topological superconductor hosting edge states with mixed chiralities. 
    \end{abstract}
	
    \maketitle
	
    \textit{Introduction}---. Dissipationless transport is long longed for by scientific community. Its early realization is brought up by the advent of superconductivity \cite{Onnes1911P,Bardeen1957PR}. 
    In a superconductor, electrons pair to form Cooper pairs, undergoing condensation below the critical temperature, which carry a nondissipative supercurrent. Another possibility is provided by a more recent achievement on the states of matters with nontrivial topology. Typical examples are the quantum Hall state \cite{Klitzing1980PRL,Thouless1982PRL,Hatsugai1993PRL} and the quantum anomalous Hall (QAH) state \cite{Haldane1988PRL,Chang2013S,Chang2023RMP}. They exhibit vanishing longitudinal resistivity and are quite robust because of their nature in topology \cite{Nakahara2003}. Then what if a conventional superconductor and the quantum Hall state are combined? The answer could be time-reversal symmetry breaking chiral topological superconductivity~\cite{Qi2010PRB}. Chiral topological superconductors~(TSCs) accommodate Majorana zero modes~\cite{Read2000PRB,Qi2011RMP,Alicea2012RPP,Beenakker2013ARCMP,Elliott2015RMP,Wilczek2009NP}, which are attractive building blocks for quantum computers~\cite{Nayak2008RMP,Karzig2017PRB,Lian2018PNAS,Zhou2019PRB}, when creating superconducting vortices. As naturally occurring chiral TSCs are rare, it is important to engineer chiral TSCs in artificial structures, where an external magnetic field or magnetization is usually considered a necessary ingredient to break time-reversal symmetry~\cite{Qi2010PRB,NadjPerge2013PRB,Braunecker2013PRL}.
	
    Very recently, the quasi-two-dimensional (2D) $ A\ce{V3Sb5} $ ($ A = $ K, Rb, and Cs) family of materials has been experimentally confirmed as a platform for various exotic quantum phenomena, including nontrivial band topology \cite{Ortiz2020PRL,Feng2021SB}, unconventional charge density wave (CDW) ordered states such as the 2-by-2 vector charge density wave (vCDW), the charge bond order (CBO), and the time-reversal symmetry breaking chiral flux phase (CFP) \cite{Feng2021PRB,Feng2021SB,Liu2022NC,Song2022SCPM}, superconductivity \cite{Ortiz2020PRL,Hao2022PRB,Jiang2022NSR}, as well as their coexistence \cite{Zheng2022N,Yu2021NC}. Moreover, although the materials themselves take three-dimensional structures, those important features are believed to originate from their 2D kagome lattice substructure \cite{Ortiz2019PRM,Ren2021PRL,Peng2021PRL,Yu2022PRL,Luo2022PRB,Kim2023NC}.
 
    Specifically, in that family of materials, the superconducting pairing symmetry is at first supported as conventional s-wave \cite{Mu2021CPL,Luo2022NC,Duan2021SCPM,Feng2022APS}, but experimental evidence of unconventional pairing soon follows up, as the absence of the Hebel-Slicheter resonant peak under pressure \cite{Zheng2022N} together with the V-shaped $ \md I/\md V $ curves from the scanning tunneling microscopy measurement \cite{Liang2021PRX,Xu2021PRL}. Additionally, inside superconducting vortices, a robust and nonsplit zero-bias conductance peak has been observed, indicating the possible presence of Majorana bound states \cite{Liang2021PRX}. 
    While relevant advancements are indeed very encouraging, it has been largely unexplored whether the coupling between superconductivity and time-reversal symmetry breaking CFP induces novel nontrivial phenomena. To be specifical, there have been so far at least three unclear issues about this family of materials:
    \begin{inparaenum}[{Issue} (I)]
		\item the undetermined superconductivity pairing symmetry;
		\item the connection between topological superconductivity and time-reversal asymmetric CFP; and
		\item the basic transport characteristics of corresponding gapless edge states, if Issue (II) is established.
    \end{inparaenum}
	
    Motivated by the above issues, in this Letter, we study the interplay between superconductivity and charge orders in a Rashba spin-orbit coupled kagome lattice by considering three possible CDW orderings and two spin-singlet (s- and $ \text{d}_{x^2-y^2} $-wave) pairing symmetries. We find that starting from the time-reversal symmetry breaking CFP state, chiral TSC phases emerge for both singlet pairings, but with distinct phase diagrams and different scenarios for chiralities of the gapless edge modes. Furthermore, because of the status of mixing chiralities, the tunneling signals plateau at different values for QAH (1) and metallic leads (\sfrac{3}{2}), respectively, for the s-wave-based TSC with $ \mathcal{N} = 1 $. After clarifying all above, we then naturally provide clues to the first issue in distinguishing between pairing symmetries. Because in the situations we have considered, there are no common TSC Chern numbers between the s-wave and the $ \text{d}_{x^2-y^2} $-wave cases.
	
\begin{figure}[htp!]
		\centering
		\includegraphics[width=.48\textwidth]{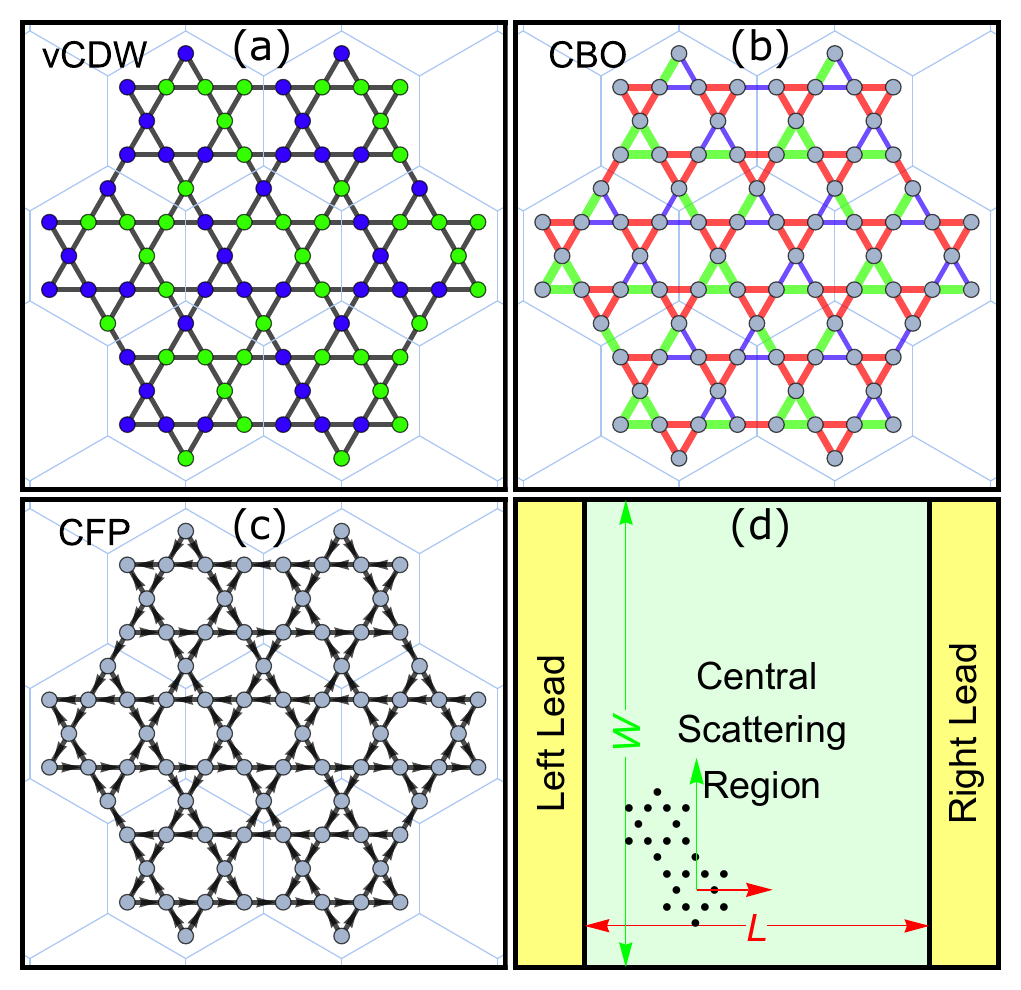}
		\caption{\label{fig1:CDWs_and_2TDev} (Color online) Panels (a)-(c): Graph representations of the three kinds of CDW states, where the light-blue-edged hexagons represent the primitive cells. In Panel (a), the blue and green nodes have negative and positive onsite energy modification to the chemical potential $ -(\mu \pm \lambda^\text{vCDW}) $, and the black edge connecting a pair of nodes represents the nearest-neighbor hopping. In Panel (b), all the onsite energies are the same, but the hoppings are modified: the red edges stands for the normal hopping $ -t $, the thicker green ones enhanced $ -(t + \lambda^\text{CBO}) $, and the thinner blue ones weakened $ -(t - \lambda^\text{CBO}) $. In Panel (c), the hopping is modified by a pure imaginary number to account for the chiral flux $ -(t \pm \mi \lambda^\text{CFP}) $, where the sign is indicated by the directions of the arrows. Panel (d): A schematic depiction of a two-terminal device to study quantum transport behavior of the topological superconducting phase discovered in this work, where an ($ L \times W $)-sized central scattering region is constructed by translating the inset minimal unit along the red and green vectors $ L $ and $ W $ times, respectively, with the atoms (with dangling bonds) on both edges trimmed.} 
    \end{figure}

    \textit{Model Hamiltonian}---. In order to achieve the purpose stated above, we build our BdG model based on the 2-by-2 CDW tight-binding Hamiltonian \cite{Feng2021PRB,Feng2021SB} with ingredients including Rashba spin-orbit coupling due to the breaking of inversion symmetry by the substrate and two kinds of superconducting pairing terms: the conventional s-wave and the unconventional $ \text{d}_{x^2-y^2} $-wave. In the second quantization form, it can be written as
    \begin{align}\label{eq:Hamiltonian}
		H_\text{BdG} &= H_0 + H_\text{CDW} + H_\text{RSOC} + H_\text{SC}, \\
		H_0 &= \sum_i c_i^\dagger (-\mu) c_i + \sum_{\braket{ij}} c_i^\dagger (-t) c_j, \notag\\
		H_\text{CDW} &= \sum_i c_i^\dagger (-\lambda_i^\text{vCDW}) c_i + \sum_{\braket{ij}} c_i^\dagger (-\lambda_{ij}^\text{CBO} - \mi \lambda_{ij}^\text{CFP}) c_j, \notag\\
		H_\text{RSOC} &= \mi \alpha \sum_{\braket{ij}} c_i^\dagger (\bm{s}\times\hat{\bm{d}}_{ij})_z c_j, \notag\\
		H_\text{SC} &= \sum_i c_{i\uparrow}^\dagger \mathit\Delta^\text{s} c_{i\downarrow}^\dagger + \sum_{\braket{ij}} c_{i\uparrow}^\dagger \mathit\Delta_{ij}^{\text{d}_{x^2-y^2}} c_{j\downarrow}^\dagger + \Hc, \notag
    \end{align}
    where $ c_i^\dagger = (c_{i\uparrow}^\dagger, c_{i\downarrow}^\dagger) $ is the electron creation operator at the site $ i $ with the spin degree of freedom included. The pristine term $ H_0 $ contains the chemical potential $ \mu $ and the nearest hopping $ t = 1 $, which we choose to be the energy unit hereafter. The second term $ H_\text{CDW} $ describes the three kinds of CDW states, among which we consider only one at a time and they can be quantitatively mapped to graphs, respectively, into Panels (a)-(c) of \autoref{fig1:CDWs_and_2TDev}. The bracket $ \braket{\cdots} $ under the summation symbol means the first nearest neighbors. The third term $ H_\text{RSOC} $ is the Rashba spin-orbit coupling measured by the parameter $ \alpha $ and $ \hat{\bm{d}}_{ij} $ is the unit vector along the hopping direction from the site $ j $ to the site $ i $, and $ \bm{s} = (s_1, s_2, s_3) $ is the spin Pauli matrix vector. The last term $ H_\text{SC} $ accounts for the spin-singlet s-wave and $ \text{d}_{x^2-y^2} $-wave superconductivity with pairing potential, respectively, the isotropic $ \mathit\Delta^\text{s} $ and the anisotropic $ \mathit\Delta_{ij}^{\text{d}_{x^2-y^2}} = \mathit\Delta^{\text{d}_{x^2-y^2}}\cos(2\phi_{ij}) $ \cite{BlackSchaffer2012PRL,BlackSchaffer2014JPCM,Wu2021PRL,Fedoseev2022PRB,Perconte2022AP}, where $ \phi_{ij} $ is the azimuthal angle of $ \hat{\bm{d}}_{ij} $ \footnotemark[1].
	
    \footnotetext[1]{See the Supplementary Materials for further details and references therein, which include Refs.~\cite{Sun2009JPCM,Zhang2017PRB,Qiao2007N,Lima2018PRB,Paz2019AQT,Fukui2005JPSJ,Yu2011PRB,Qiao2010PRB,Jiang2012PRB,Zeng2022PRB}.}
	
\begin{figure}[htp!]
		\centering
		\includegraphics[width=.45\textwidth]{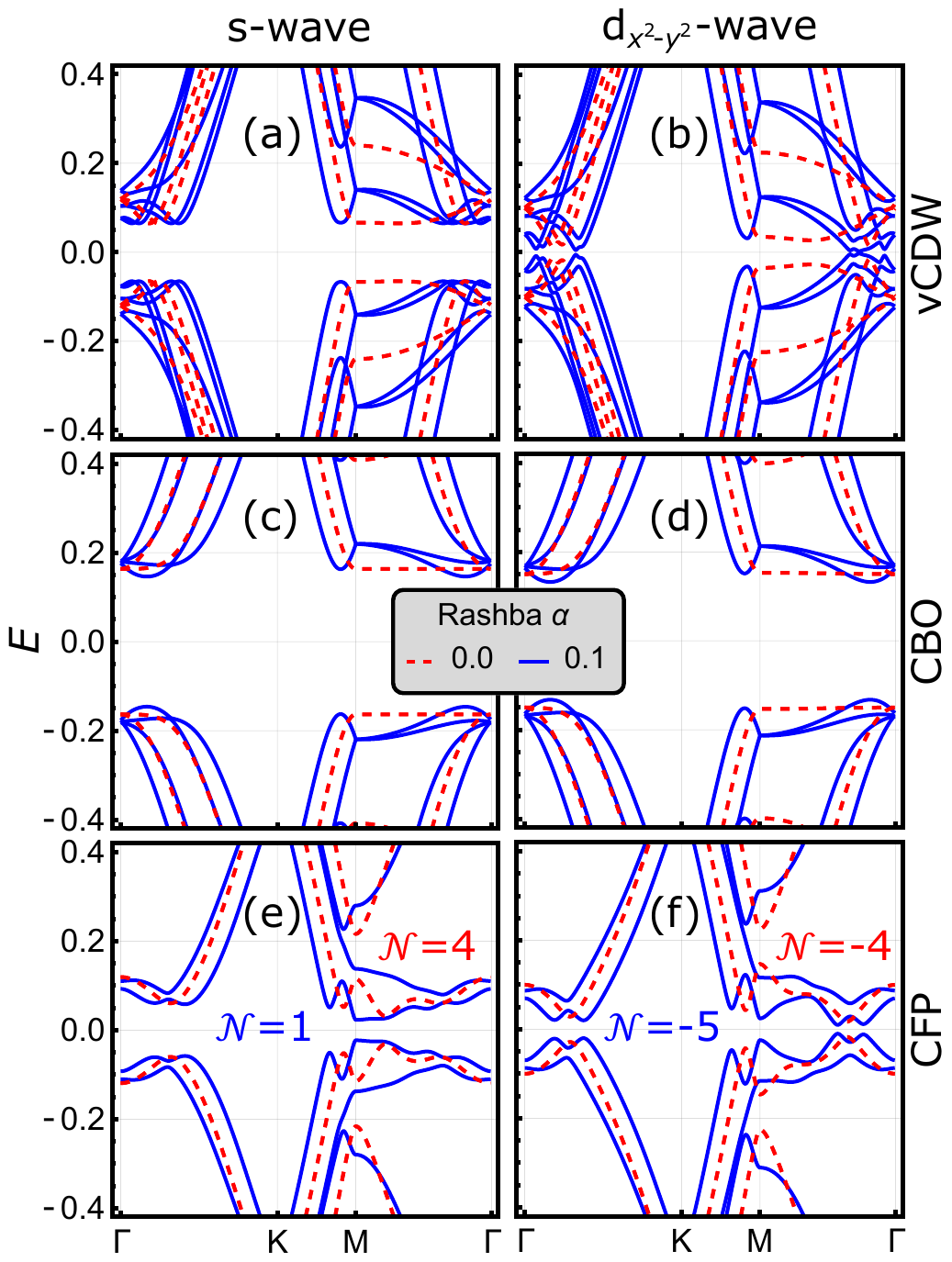}
		\caption{\label{fig2:bulk_band_structures} (Color online) Bulk band structures by combination of three CDW phases and two superconducting pairings. Panels (a)-(d): Without breaking the time reversal symmetry, a phase with a nonvanishing BdG Chern number cannot be found from the first two CDW phases. Panels (e)-(f): Topologically nontrivial superconducting phases emerge from the last CDW order. Other parameters are $ \mu = 0.1, \lambda^\text{vCDW/CBO/CFP} = 0.25, \mathit\Delta^\text{s} = 0.065 $, and $ \mathit\Delta^{\text{d}_{x^2-y^2}} = 0.15 $. All energy dimensioned quantities are measured in $ t $.}
    \end{figure}

    \textit{Topological Superconducting Phases}---.
    The model \autoref{eq:Hamiltonian} provides us the opportunity to study the quantum phases from the combination of the three CDW orderings and the two superconducting parings. As shown by the dashed red bands in Panels (a)-(f) of \autoref{fig2:bulk_band_structures}, when the system is free of Rashba SOC effect ($ \alpha = 0.0 $), with a proper set of parameters, insulating states can be obtained across all the situations under consideration. And the two superconducting pairings do not render the band structures with much difference for each CDW state. Because the first two CDW (vCDW and CBO) states respect time-reversal symmetry, a nonzero BdG Chern number is not expectable. On the contrary, as the pure CFP carries a nonvanishing Chern number $ \mathcal{C} = 2 $ (with spin degeneracy), a corresponding BdG model in the same phase reasonably doubles that value. A finite Rashba SOC ($ \alpha = 0.1 $) lifts degeneracy at most of the lattice momentum points. Although it makes the band structures more complicated, for vCDW and CBO states, a nontrivial phase is still missing. However, the Rashba effect induces topological phase transitions for the CFP order, and odd integer BdG Chern numbers can be acquired $ \mathcal{N} = 1 $ for s-wave and $ \mathcal{N} = -5 $ for $ \text{d}_{x^2-y^2} $-wave pairing, respectively.
	
\begin{figure}[htp!]
		\centering
		\includegraphics[width=.45\textwidth]{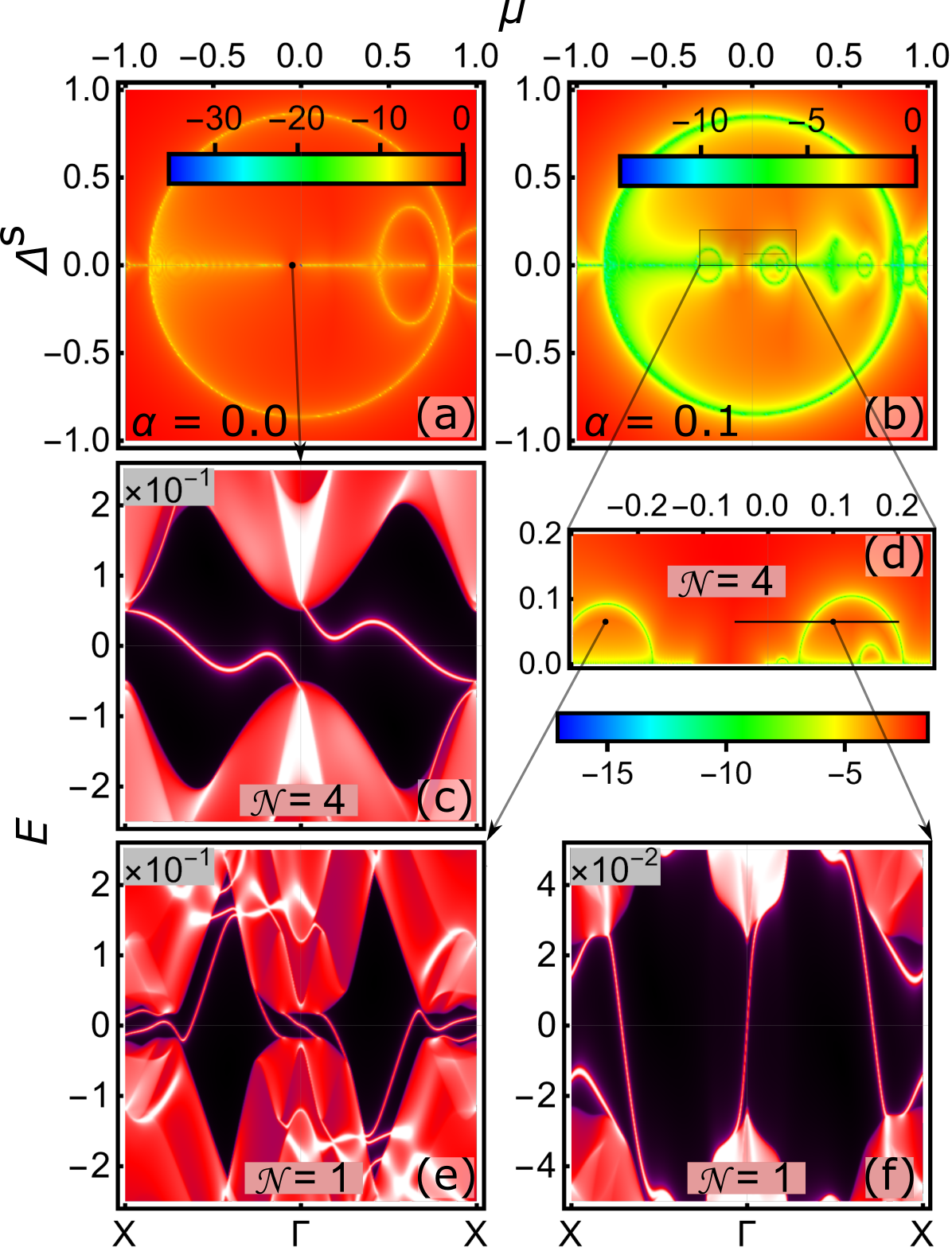}
		\caption{\label{fig3:phase_diagrams_and_spectral_data_s-wave} (Color online) Phase diagrams for CFP with s-wave superconducting pairings and relevant gapless edge states for TSC. Panel (a): Without a Rashba effect, most region occupied by the $ \mathcal{N} = 4 $ phase, whose corresponding edge states are shown in Panel (c), with spin degeneracy. Panel (b): With a Rashba effect, TSC phases appear, where the region of interest is zoomed in in Panel (d), and the edge states of two typical TSC phases are respectively shown in Panels (e) and (f). The horizontal line segment in Panel (d) indicates the parameter samplings for transport study part of this work shown in \autoref{fig5:trans_coeffs}.}
    \end{figure}

    A more thorough perspective can be obtained by investigating the topological phase diagram in the $ \mathit\Delta $-$ \mu $ space, with color encoding the logarithm of band gap. We first check the case of s-wave in \autoref{fig3:phase_diagrams_and_spectral_data_s-wave}. When Rashba effect does not exist ($ \alpha = 0.0 $), the BdG model in \autoref{eq:Hamiltonian} just simply doubles the chiral flux phase ($ \mathcal{N} = 4 = 2\mathcal{C} $) in most area of the phase space [see \autoref{fig3:phase_diagrams_and_spectral_data_s-wave}(a)] and its corresponding spectral function data is presented in \autoref{fig3:phase_diagrams_and_spectral_data_s-wave}(c), where the gapless edge state chirality concurs with the BdG Chern number, with the spin degeneracy understood. There is a major yellow phase boundary (seemly a circle), beyond which it is a totally trivial phase. And the size of that phase border is directly related to the value of $ \lambda_\text{CFP} $. When a finite Rashba effect plays its role ($ \alpha = 0.1 $), some new regions with various shapes and corresponding boundaries appear accordingly from the vanishing-Rashba framework [see \autoref{fig3:phase_diagrams_and_spectral_data_s-wave}(b) and its relevant part zoomed-in in \autoref{fig3:phase_diagrams_and_spectral_data_s-wave}(d)]. At least two adiabatically unconnected TSC phases both with $ \mathcal{N} = 1 $ are among them, whose corresponding scenarios of gapless modes are displayed, respectively, in the Panels (e) and (f) of \autoref{fig3:phase_diagrams_and_spectral_data_s-wave}. One can see that, although possessing the same Chern number, the situations of gapless modes of the two phases are very different from each other. This CFP-based TSC phase diagram possesses more complexity than its square-lattice counterpart \cite{Qi2010PRB}. The situations for the case of $ \text{d}_{x^2-y^2} $-wave bear many similarities, where TSC phases appear with the first three negative odd BdG Chern numbers \footnotemark[1]. All of the TSC phases shown here have the same characteristics that the total number of edge states is larger than the corresponding Chern number, but none of the net chirality is violated.

\begin{figure}[htp!]
		\centering
		\includegraphics[width=.49\textwidth]{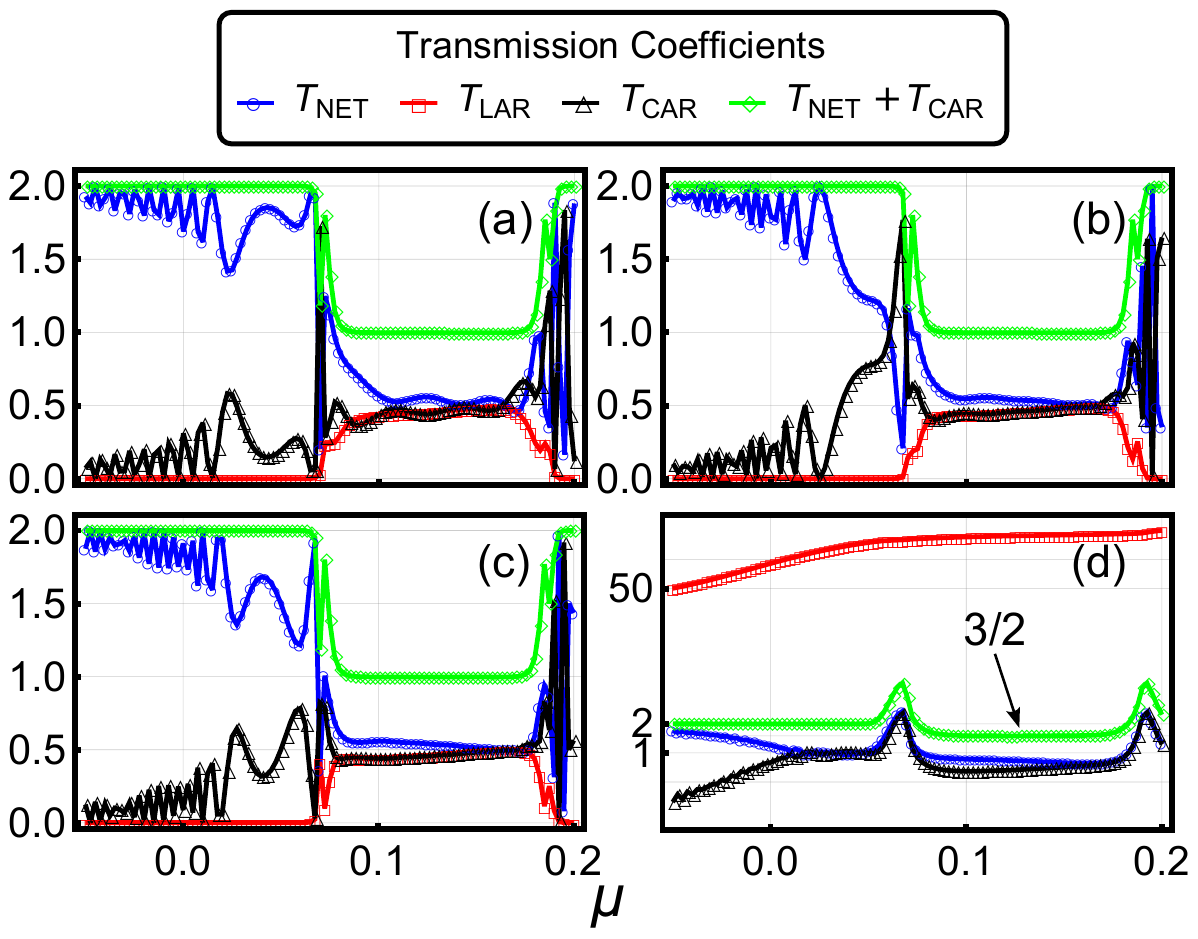}
		\caption{\label{fig5:trans_coeffs} (Color online) TSC-related transport coefficients for a two-terminal device with an ($ L \times W $)-sized sheet as its central scattering region. For all panels, the sheet width is fixed at $ W = 80 $, the superconducting pairing potential is $ \mathit{\Delta}^\text{s} = 0.065 $, and $ E = 0^+ $. Panels (a)-(c): Chern insulator leads with the sheet length $ L = 40, 50 $, and $ 60 $, respectively. Panel (d): Metallic leads with the length $ L = 60 $. Though the curves behave in a more complex manner, tunneling plateau signals (1 and \sfrac{3}{2}) are still accessible, despite the size of sheet or the category of leads. (Abbr.: NET = normal electron tunneling, LAR = local Andreev reflection, and CAR = crossed Andreev reflection; $ \mu $ in unit of $ t $)}
	\end{figure}

	\textit{Transport Properties of The Gapless Edge States}---.	
	Now that those TSC phases whose gapless edge modes possess mixed chiralities have been figured out, more can be elucidated by studying their quantum transport behaviors. To start with, we focus on the s-wave case with $ \mathcal{N} = 1 $, and choose the area where the values of parameters are not very large, where we indicate a straight line path [see \autoref{fig3:phase_diagrams_and_spectral_data_s-wave}(d)] with a fixed pairing potential but a varying chemical potential that walks through the trivial and nontrivial superconducting states. Also in view of that a junction with Chern insulating leads and its derivatives have been shown to have a close relationship with Majorana braiding operations \cite{Lian2018PNAS}, and metallic leads are the most experimentally accessible, we build up corresponding two-terminal devices containing a finite sheet with sufficient sizes as its central scattering region, as schematized in \autoref{fig1:CDWs_and_2TDev}(d). The Chern insulator leads are in the state shown in \autoref{fig3:phase_diagrams_and_spectral_data_s-wave}(c) and the metallic leads are those with merely the $ t $-term in \autoref{eq:Hamiltonian} and all other parameters are set to be zero. We then find out the coefficients during the processes of the normal electron tunneling (NET), the crossed Andreev reflection (CAR), and the local Andreev reflection (LAR) through the lattice Green's function method with Landauer-Büttiker formalism, with the adaptive partition of the central scattering region used \footnotemark[1]. The major results are presented in \autoref{fig5:trans_coeffs} for a central scattering region with a size of $ L \times W $, where the width along the $ y $-direction is fixed at a sufficiently large value as $ W = 80 $, so that the chiral edge states on the upper and the lower edges hardly mix, and the length along the $ x $-direction is chosen as $ L = 40, 50 $, and $ 60 $. The two leads each have an appropriate semi-infinite lattice translational symmetry along the $ x $-direction \footnotemark[2].
	
	\footnotetext[2]{See \autoref{fig1:CDWs_and_2TDev}(d) and its caption for the exact information for the size.}
	
	With the superconducting pairing potential fixed at $ \mathit{\Delta}^\text{s} = 0.065 $ and $ E = 0^+ $, each coefficient varies with respect to the change of the chemical potential $ \mu \in [-0.05, 0.20] $, corresponding to the selected path shown in the phase diagram in \autoref{fig3:phase_diagrams_and_spectral_data_s-wave}(d), where one can see that around $ \mu_\text{c} \approx 0.07 $ there is a topological phase transition. Therefore when the chemical potential is small and the central scattering region is in $ \mathcal{N} = 4 $ phase, none of the coefficients exceed two [\autoref{fig5:trans_coeffs}(a)-(c)], because there are only two injecting electron chiral edge states when the leads are all in a QAH effect with $ \mathcal{C} = 2 $. Specifically, the process of LAR is totally suppressed, and that for NET oscillates under the value of two, so does the other tunneling process of CAR, so that even though each one of them individually vibrates but together contribute to a plateau signal of two.
	
	When the chemical potential continues increasing to pass over the critical value ($ \mu_\text{c} \approx 0.07 $), a topological phase transition occurs and now the finite flake enters the TSC phase with $ \mathcal{N} = 1 $. Then one can see that the amplitudes of coefficients of the both tunneling processes (the NET and the CAR) are reduced, however the LAR coefficient obtains its finite values and largely the same pace with that of the CAR. This coincidence behavior between the two kinds of transport processes has been reported previously \cite{Li2020PRB}, but now neither of them is constant with respect to the chemical potential. On the other hand, the tunneling processes together still provide plateau information at a unit value, which can be understood, as the two incoming electron chiral edge states invoke the two Majorana chiral edge states with the same moving direction in the central scattering region, so that the plateau value drops a half in magnitude after the topological phase transition happens, which moreover is insensitive to the length of the finite sheet [\autoref{fig5:trans_coeffs}(a)-(c)]. Finally, when the chemical potential gets close to 0.2, the tunneling plateau rises up to two again, and simultaneously the LAR is vanishing, because the system crosses the phase border the second time. We notice that in the course of phase transition (within two small windows of $ \mu $ around 0.07 and 0.18), the tunneling coefficients experience drastic oscillations. What happens there is yet to be studied but beyond the scope of present work.
	
	With other conditions maintained, now we change the leads to a normal metallic state so that the incoming states not only increase in number but also include bulk wavefunctions, and the corresponding transport behavior is shown in \autoref{fig5:trans_coeffs}(d). One can see that in this case, a large number of inscattering electrons are responded by the process for reflection of holes. And the tunneling processes provide a plateau of two, same as the case with Chern insulating leads [\autoref{fig5:trans_coeffs}(a)-(c)] when $ \mathcal{N} = 4 $, and there is an obvious widow of chemical potential where the NET and the CAR approximately have the same coefficient value. When $ \mathcal{N} = 1 $, however, the plateau changes from previous unity to exhibiting an extra half. This is also comprehensible because now the Majorana chiral edge state with the opposite chirality but localized at the opposite edge is also invoked, so there are totally three Majorana chiral edge states participating in the transport processes in this case.
	
	\textit{Summary and Discussion}---.
	To summarize, we establish the connection between topological superconductivity and time-reversal asymmetric CFP state and then confirm a series of topological superconducting phases in the CDW-ordered kagome systems. However, unlike the conventional cases, those new phases are found that edge states highly possibly carry mixed chiralities, e.g., a unit BdG Chern number can support three chiral edge states, yet without violating the chirality constraint. So those phases are expected to provide multiple transport channels and more complexity possibly would be introduced, which is verified in the result of transport coefficients for a two-terminal device. Taken as an example, for the s-wave case with $ \mathcal{N} = 1 $, when Chern insulator is used as the leads, the tunneling processes together lead to a unity plateau, and the same kind of plateau becomes $ \sfrac{3}{2} $ when there are an abundant of incoming metallic states. Our work reveals the basic transport properties of topological superconductivity when its gapless edge states carry mixed chiralities, which is different from both the pure chiral and the helical topological superconductors. Because the s-wave and the $ \text{d}_{x^2-y^2} $-wave correspond to different chiral TSC phases, these findings not only provide a clue to distinguish between the superconductivity pairing symmetries but also offer a promising avenue for exploring novel quantum phenomena in the kagome materials $A$V$_3$Sb$_5$.

    \textit{Acknowledgement}---.
    This work was supported by the National Natural Science Foundation of China (NSFC, Grants No.~12247181, No.~12222402, No. 12074108, No.~11974062, and No.~12147102).
	
    \bibliography{CFP_TSC_transport}
	\bibliographystyle{apsrev4-2} 
\end{document}